\documentclass[prl,aps,floatfix,showpacs,twocolumn]{revtex4}
\usepackage{dcolumn}
\usepackage{bm}
\usepackage{color}
\usepackage{graphicx}
\usepackage{epsf}
\usepackage{pstricks}
\begin{document}
\title{Chiral effective field theory predictions for muon capture on deuteron
and $^3$He}
\author{L.E.\ Marcucci$^{\, {\rm a,b}}$, 
A.\ Kievsky$^{\,{\rm b}}$, 
S.\ Rosati$^{\, {\rm a}}$, 
R.\ Schiavilla$^{\,{\rm c,d}}$, and M.\ Viviani$^{\,{\rm b}}$}
\affiliation{
$^{\,{\rm a}}$\mbox{Department of Physics, University of Pisa, 56127 Pisa, Italy}\\
$^{\,{\rm b}}$\mbox{INFN-Pisa, 56127 Pisa, Italy}\\
$^{\,{\rm c}}$\mbox{Department of Physics, Old Dominion University, Norfolk, VA 23529, USA}\\
$^{\,{\rm d}}$\mbox{Jefferson Lab, Newport News, VA 23606}\\
}

\date{\today}

\begin{abstract}
The muon-capture reactions $^2$H($\mu^-,\nu_\mu$)$nn$ and
$^3$He($\mu^-,\nu_\mu$)$^3$H are studied with nuclear 
potentials and charge-changing weak currents, derived in chiral effective
field theory.  The low-energy constants (LEC's) $c_D$ and $c_E$, present in the
three-nucleon potential and ($c_D$) axial-vector current, are constrained
to reproduce the $A=3$ binding energies and the triton Gamow-Teller matrix element.
The vector weak current is related to the isovector component of the 
electromagnetic current via the conserved-vector-current constraint, and the two LEC's 
entering the contact terms in the latter are constrained to reproduce the
$A=3$ magnetic moments.  The muon capture rates on deuteron and $^3$He
are predicted to be $399\pm 3$ sec$^{-1}$ and $1494\pm 21$ sec$^{-1}$,
respectively. The spread accounts for the cutoff sensitivity as well
as uncertainties in the LEC's and electroweak radiative corrections.
By comparing the calculated and precisely measured rates on $^3$He,
a value for the induced pseudoscalar form factor is obtained
in good agreement with the chiral perturbation theory prediction.
\end{abstract}


\index{}\maketitle
When negative muons pass through matter, they can be captured into high-lying
atomic orbitals.  They then quickly cascade down into the 1$S$ orbit, where two
competing processes occur: one is ordinary decay
$\mu^-\rightarrow e^-\,\overline{\nu}_e\, \nu_\mu$, and the other is (weak)
capture by the nucleus $\mu^-\, A(Z,N)\rightarrow \nu_\mu \, A(Z-1,N+1)$.
Apart from tiny corrections due to bound-state effects (chief among which
is time-dilation)~\cite{Cza00}, the decay rate is essentially the same as for
a free muon and, in light nuclei, is much larger than the rate for capture.  The latter
proceeds predominantly through the basic process $p\, \mu^-\rightarrow n\, \nu_\mu$ 
induced by the exchange of a $W^+$ boson, and its rate, which would naively
be expected to scale with the number $Z$ of protons in the nucleus, is 
enhanced by an additional flux factor of $Z^{\,3}$, originating from the square of
the atomic wave function (w.f.) evaluated at the origin~\cite{Pri59}.  Thus capture,
with a rate 
proportional to $Z^4$,
dominates decay at large $Z$.

Muon capture on hydrogen is, in principle, best suited to obtain
information on the nucleon matrix element of the charge-changing
quark current $\overline{d}\gamma^\mu(1-\gamma_5)u$, responsible
for the process $p\, \mu^-\rightarrow n\, \nu_\mu$. Ignoring
contributions from second-class currents~\cite{Wei58} for which
there is presently no firm experimental evidence~\cite{Sev06}, it is parametrized
in terms of four form factors (f.f.'s): two, $F_1(q^2)$ and $F_2(q^2)$,
from the polar-vector component of the weak current are related to the
(isovector) electromagnetic (EM) form factors of the nucleon by the
conserved-vector-current (CVC) constraint; two, the axial and
induced pseudoscalar f.f.'s $G_A(q^2)$ and $G_{PS}(q^2)$, go along
with the axial-vector part of the weak current.  The $F_1(q^2)$ and $F_2(q^2)$
f.f.'s are well known over a wide range of momentum transfers $q^2$
from elastic electron scattering data on the nucleon~\cite{FF}.  
The f.f.\ $G_A(q^2)$ is also quite well known: its value at vanishing $q^2$, $g_A=1.2695 (29)$, is
from neutron $\beta$-decay~\cite{PDG}, while its $q^2$-dependence
is parametrized as $G_A(q^2)=g_A/( 1-q^2/\Lambda_A^2)^2$, with
$\Lambda_A= 1$ GeV from an analysis of pion electro-production data~\cite{Ama79}
and direct measurements of the reaction $p\, \nu_\mu\rightarrow n\, \mu^+$~\cite{Kit83}.

Of the four f.f.'s, the induced pseudoscalar $G_{PS}(q^2)$ is the least known.
The MuCap collaboration at PSI has recently reported a precise measurement of
the rate for capture on hydrogen in the 1$S$ singlet hyperfine state:
$\Gamma(^1{\rm H})|_{\rm singlet}=
725.0 \pm 13.7 ({\rm stat}) \pm 10.7 ({\rm syst})$ sec$^{-1}$~\cite{MuCap}.
Based on this value, an indirect \lq\lq experimental\rq\rq determination of $G_{PS}$ at the
momentum transfer $q_0^2=-0.88\, m_\mu^2$ relevant for $\mu^-$
capture on hydrogen, $G^{\,{\rm EXP}}_{PS}(q_0^2)=7.3\pm 1.2$, has been 
obtained by using
for the remaining f.f.'s the values discussed above and by evaluating
electroweak radiative corrections~\cite{Cza07}.  The latter lead to a 2.8\% increase
in the rate on hydrogen, and are crucial for bringing $G^{\,{\rm EXP}}_{PS}$
within less than 1$\sigma$ of the most recent theoretical
prediction, $G^{\rm TH}_{PS}(q_0^2)=8.2 \pm 0.2$~\cite{GPth},
obtained in chiral perturbation theory ($\chi$PT).
For a recent and comprehensive review of theoretical and experimental
efforts to determine $G_{PS}(q^2)$ see Refs.~\cite{Gor04,Kam10}.

In the present letter, we focus on the reactions $^2$H($\mu^-,\nu_\mu$)$nn$
and $^3$He($\mu^-,\nu_\mu$)$^3$H, hereafter referred to as $\mu$--2 and
$\mu$--3, respectively.  There are a couple of reasons for undertaking this
study now: (i) the forthcoming measurement of
the $\mu$--2 rate $\Gamma(^2{\rm H})$ in the doublet hyperfine state
by the MuSun collaboration at PSI with a projected
1\% precision ~\cite{And07,Kam10}.  This and the already available,
and remarkably precise, measurement of the $\mu$--3 rate,
$\Gamma(^3{\rm He})=1496\pm 4$ sec$^{-1}$~\cite{Ack98},
will make it possible to put tight constraints on $G^{\rm EXP}_{PS}(q^2)$ and to
test the $\chi$PT prediction for this f.f. far more sharply than up to now.
(ii) A number of low-energy weak processes of
astrophysical interest, such as the weak captures on proton and $^3$He,
and neutrino reactions on light nuclei, are not accessible experimentally.
In order to have some level of confidence
in the reliability of their cross section estimates, it becomes crucial
to study, within the same theoretical framework, related electroweak reactions, whose
rates are known experimentally, like muon captures~\cite{SFII}.

Theoretical work on the $\mu$--2 and $\mu$--3 reactions is quite extensive
(see Refs.~\cite{Mea01,Gor04,Kam10,Mar11}).
So far, calculations have been performed within two different schemes:
the ``standard
nuclear physics approach'' (SNPA) and the approach known as ``hybrid'' chiral
effective
field theory ($\chi$EFT).  In SNPA, Hamiltonians based
on conventional two-nucleon (NN)
and three-nucleon (NNN) potentials are used to calculate the nuclear w.f.'s,
and the weak transition operator includes, beyond the one-body
contribution (the impulse approximation---IA)
associated with the basic process $p\, \mu^-\rightarrow n\, \nu_\mu$,
meson-exchange currents as well as
currents arising from the excitation of $\Delta$-isobar degrees of freedom~\cite{Mar00}.
In  the hybrid $\chi$EFT approach,
the weak operators are derived in $\chi$EFT, but
their matrix elements  are evaluated between w.f.'s
obtained from conventional potentials.  Typically, the SNPA and
hybrid $\chi$EFT predictions are in good agreement with each other.  For example,
for the $\mu$--2 rate, the
SNPA calculation of Ref.~\cite{Mar11} gives 391 sec$^{-1}$, to be compared
with the hybrid $\chi$EFT studies of Refs.~\cite{And02} and~\cite{Mar11},
which report 386 sec$^{-1}$ and $393 \pm 1$ sec$^{-1}$, respectively.
The differences between Refs.~\cite{And02} and~\cite{Mar11}
are due to contributions of loop corrections and contact terms in the
vector part of the weak current, which were neglected in Ref.~\cite{And02}.
For the $\mu$--3 rate, the SNPA calculation of Ref.~\cite{Mar11} gives 
1486 sec$^{-1}$, while the hybrid $\chi$EFT studies of Refs.~\cite{Gaz08} 
and~\cite{Mar11} report, respectively, $1499\pm 16$ sec$^{-1}$ and 
$1484 \pm 4$ sec$^{-1}$. Here, the differences
between Refs.~\cite{Gaz08} and~\cite{Mar11} arise mostly from
the  inclusion in Ref.~\cite{Gaz08} of vacuum polarization effects on the 
muon bound state w.f.~\cite{Cza07}---these would increase the SNPA and
hybrid $\chi$EFT results of Ref.~\cite{Mar11}
quoted above for the $\mu$--3 rate to, respectively, 1496 sec$^{-1}$ and 
$1494 \pm 4$ sec$^{-1}$.

One of the objectives of the present work is to carry out a
$\chi$EFT calculation of the $\mu$--2 and $\mu$--3 rates.  Chiral
EFT is a formulation of quantum chromodynamics (QCD)
in terms of effective degrees of freedom suitable for low-energy
nuclear physics: pions and nucleons.  The symmetries of QCD, in
particular its (spontaneously broken) chiral symmetry, severely
restrict the form of the interactions of nucleons and pions
among themselves and with external electroweak fields, and
make it possible to expand the Lagrangian describing
these interactions in powers of $Q/\Lambda_\chi$, where
$Q$ is pion momentum and $\Lambda_\chi \sim 700$ MeV
is the chiral-symmetry-breaking scale.  As a consequence,
classes of Lagrangians emerge, each characterized by
a given power of $Q/\Lambda_\chi$ and each involving
a certain number of unknown coefficients, so called
low-energy constants (LEC's).  While these LEC's could in principle
be determined by theory (for instance, in lattice QCD calculations),
they are in practice constrained by fits to experimental data.
Some of them (for example, $g_A$ and the pion decay amplitude $F_\pi$)
characterize the coupling (at lowest order) of pions to nucleons and,
in particular, the strength of one- and two-pion-exchange
terms (denoted, respectively, OPE and TPE)
in the NN potential~\cite{Ent03,Mac11}, {\it i.e.}~its long-range
components.  Some of the other LEC's multiply NN (and
multinucleon) contact interactions, and therefore encode short-range
physics, which in a meson-exchange picture would, for example,
be associated with vector-meson exchanges or excitation
of baryon resonances, like the $\Delta$ isobar.

The NN potential has been derived up to order $(Q/\Lambda_\chi)^4$
in the chiral expansion~\cite{Ent03,Mac11}.  It consists of OPE and
TPE with interaction vertices from leading, next-to-leading, and next-to-next-to-leading
$\pi$N chiral Lagrangians, 
and of contact terms.  The LEC's have been constrained by accurate fits
to the NN scattering database at energies below the pion
production threshold (see Ref.~\cite{Mac11} for a review).
The NNN potential, which first contributes at order $(Q/\Lambda_\chi)^3$,
includes $S$- and $P$-wave TPE---its $P$-wave piece is the familiar
Fujita-Miyazawa NNN potential---a
OPE plus NN contact term with LEC $c_D$ and a
NNN contact terms with LEC $c_E$.

The vector and axial pieces of the weak
current have been derived up to order $Q/\Lambda_\chi$
in, respectively, Refs.~\cite{Par96,Pastore09} and~\cite{Par03}.
The one-body operators are the same as
those obtained in the SNPA by retaining, in the expansion of the covariant
single-nucleon four-current, corrections up to order $(v/c)^2$ relative to the
leading-order term~\cite{Mar00}.  Two-body operators in the
axial current (charge) first enter at order $(Q/\Lambda_\chi)^0$
[$(Q/\Lambda_\chi)^{-1}$], and are suppressed, in the power
counting, by $(Q/\Lambda_\chi)^3$ [$Q/\Lambda_\chi$]
relative to the one-body term of order $(Q/\Lambda_\chi)^{-3}$
[$(Q/\Lambda_\chi)^{-2}$].  In the axial current,
these terms include a OPE contribution, involving the
known LEC's $c_3$ and $c_4$ (determined by fits to the NN
data~\cite{Mac11}), and
one contact current, whose strength is parametrized
by the LEC $d_R$ (see below).  In the axial charge, only OPE
contributes, and the associated operator is 
proportional to $g_A/F_\pi^2$.  
One-loop corrections to the axial charge and
current from TPE, which enter at $Q/\Lambda_\chi$ and are therefore strongly
suppressed relative to the leading-order one-body terms,
are ignored, since their contributions are expected to be tiny.

The vector weak current is related (via the CVC constraint) to
the EM current, which includes, up to order $Q/\Lambda_\chi$,
OPE and TPE ({\it i.e.}, one-loop corrections), as well as
isoscalar and isovector contact terms, whose strengths are
parametrized by the LEC's denoted, respectively,
as $g_{4S}$ and $g_{4V}$ in the following~\cite{Par96,Par03}.
It has been shown~\cite{Pastore09} that such a current
satisfies the continuity equation with the NN potential
at order $(Q/\Lambda_\chi)^2$.  In this regard, we note
that the construction of a conserved current with the $(Q/\Lambda_\chi)^4$
NN potential used here would require the inclusion of
terms up to order $(Q/\Lambda_\chi)^3$, {\it i.e.}, two-loop
corrections.  This is a daunting task, well beyond
the present state of the art.  In a more speculative vein,
it is also not obvious that such a theory could be made predictive, given the presumably
large number of contact terms with unknown LEC's that it would entail.

Finally, we notice that potentials and currents have power-law behavior
for large momenta, and need to be regularized.  This is accomplished
in practice by introducing a momentum-cutoff function.  In the
present work, the cutoff $\Lambda$ is taken to be 500 MeV and 600 MeV.

\begin{figure}[t]
\includegraphics[width=2.5in,height=1.5in]{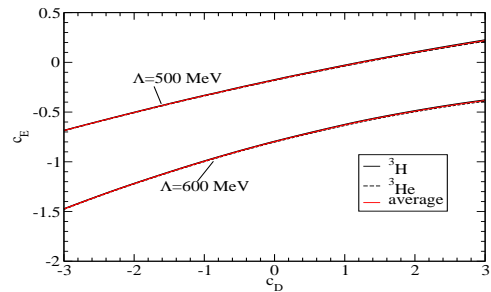}
\vspace*{0.1cm}
\caption{(Color online) $c_D$-$c_E$ trajectories fitted to
reproduce the experimental $^3$H and $^3$He BE's.  See text for explanation.}
\label{fig:cdce}
\end{figure}

\begin{figure}[t]
\vspace*{0.5cm}
\includegraphics[width=2.5in,height=1.5in]{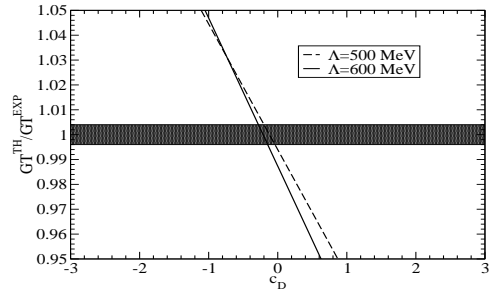}
\vspace*{0.1cm}
\caption{The ratio GT$^{\rm TH}$/GT$^{\rm EXP}$ as function of the LEC $c_D$.}
\label{fig:gt}
\end{figure}

We now turn our attention to the determination
of the LEC's $d_R$, $c_D$, $c_E$, $g_{4S}$, and $g_{4V}$.
In the past, $c_D$ and $c_E$ were fixed by fitting
the triton binding energy (BE) together with an additional 
strong-interaction observable, such as the $nd$ doublet
scattering length $^2a_{nd}$ or $^4$He BE.
However, this led to significant uncertainties, due to
strong correlations between these observables~\cite{Kie10}.
As the authors of Ref.~\cite{Gar-Gaz} have observed,
the LEC's $d_R$ and $c_D$ are related to each other via
$d_R=\frac{M_N}{\Lambda_\chi\, g_A}c_D
+\frac{1}{3}M_N(c_3+2\, c_4)+ \frac{1}{6}$ ($M_N$ is the nucleon mass), and
therefore one can fix $c_D$ (or $d_R$) and $c_E$
by fitting the triton BE and half-life (specifically, the
Gamow-Teller matrix element).  Thus, we proceed as follows.
The $^3$H and $^3$He ground state w.f.'s are calculated with
the hyperspherical-harmonics method (see Ref.~\cite{Kie08} for a review)
using the chiral NN+NNN potentials of Refs.~\cite{Ent03,Mac11,Nav07} for 
$\Lambda =$ 500 and 600 MeV. The corresponding set of
LEC's $\{c_D,c_E\}$ is determined by fitting the $A=3$ experimental
BE's, BE($^3$H)=8.475 MeV and BE($^3$He)=7.725 MeV,
corrected for small contributions (+7 keV in $^3$H and --7 keV in $^3$He)
due to the $n$-$p$ mass difference~\cite{Nog03}, since this
effect is neglected in the present calculations.  We then span the range
$c_D\in[-3,3]$, and, in correspondence
to each $c_D$ in this range, determine $c_E$ so as to reproduce
either BE($^3$H) or BE($^3$He).  The resulting trajectories are 
shown in Fig.~\ref{fig:cdce}, and are nearly indistinguishable.
Their average, shown by the red lines in Fig.~\ref{fig:cdce},
leads to $A=3$ BE's within 10 keV of the experimental values above.
Then, for each set of $\{c_D,c_E\}$, the triton and
$^3$He w.f.'s are calculated and, using the
$\chi$EFT weak axial current discussed above, the Gamow-Teller
matrix element of tritium $\beta$-decay (GT$^{\rm TH}$) is
determined.  The ratio GT$^{\rm TH}$/GT$^{\rm EXP}$
is shown in Fig.~\ref{fig:gt}, for both values of the cutoff $\Lambda$.
We have used GT$^{\rm EXP}=0.955 \pm 0.004$, as obtained in Ref.~\cite{Mar11},
except that we have conservatively doubled the error, represented by
the shadowed band in the figure.  The range of $c_D$ values, for which
${\rm GT}^{\rm TH}={\rm GT}^{\rm EXP}$ within the experimental
error, are $[-0.20, -0.04]$ for $\Lambda=500$ MeV,
and $[-0.32, -0.19]$ for $\Lambda=600$ MeV.
The corresponding ranges for $c_E$ are $[-0.208, -0.184]$
and  $[-0.857, -0.833]$, respectively.  
We note that, for each pair of $\{c_D,c_E\}$ in the selected range,
the scattering length $^2a_{nd}$
is calculated to be $^2a_{nd}=0.666\pm 0.001$ fm for
$\Lambda=500$ MeV and $^2a_{nd}=0.696\pm 0.001$ fm for
$\Lambda=600$ MeV, which should be compared with
$^2a_{nd}=0.675$ fm~\cite{Kie08}, obtained with $\Lambda=500$ MeV
and $\{c_D,c_E\}=\{1.0,-0.029\}$, as originally set in Ref.~\cite{Nav07}.
The most recent experimental determination gives
$^2a_{nd}=0.645\pm 0.010$ fm~\cite{Sch03}.

For the minimum and maximum values of $\{ c_D , c_E \}$
in the selected range, {\it i.e.}, $\{ -0.20 , -0.208 \}$ and $\{ -0.04 , -0.184 \}$
for $\Lambda=500 $ MeV, and $\{ -0.32 , -0.857 \}$ and $\{ -0.19 , -0.833 \}$
for $\Lambda=600 $ MeV, we have determined the isoscalar and isovector
LEC's, $g_{4S}$ and $g_{4V}$, entering the NN
contact terms of the EM current by reproducing the $A=3$
magnetic moments.  These LEC's are listed in Table~\ref{tab:LECs}.
\begin{table}[bthp]
\caption{The LEC's $g_{4S}$ and $g_{4V}$ associated with the
isoscalar and isovector NN contact terms in the EM
current for
$\Lambda=500$ and 600 MeV.  See text for explanation.}
\begin{tabular}{cccc}
\hline
  &  $\{ c_D , c_E \}$ & $g_{4S}$ & $g_{4V}$ \\
\hline
$\Lambda$=500 MeV  & $\{ -0.20 , -0.208 \}$ & $0.207\pm 0.007$ 
& $0.765\pm 0.004$ \\
                   & $\{ -0.04 , -0.184 \}$ & $0.200\pm 0.007$ 
& $0.771\pm 0.004$ \\
\hline
$\Lambda$=600 MeV  & $\{ -0.32 , -0.857 \}$ & $0.146\pm 0.008$ 
& $0.585\pm 0.004$ \\
                   & $\{ -0.19 , -0.833 \}$ & $0.145\pm 0.008$ 
& $0.590\pm 0.004$ \\
\hline
\end{tabular}
\label{tab:LECs}
\end{table}

Having fully constrained the NNN potential and weak current,
we present in Table~\ref{tab:res} the 
$\chi$EFT predictions for the $\mu$--2 
and $\mu$--3 rates, $\Gamma(^2{\rm H})$ and $\Gamma(^3{\rm He})$.  For $\Gamma(^2{\rm H})$, we also show the individual contributions
of $nn$ channels with total angular momentum $J\leq 2$ ($^1S_0$, $^3P_0$, $^3P_1$,
$^3P_2$, $^1D_2$ and $^3F_2$).  Higher partial waves are known to contribute less
than 0.5 \% to $\Gamma(^2{\rm H})$~\cite{Mar11}.  The one-body (IA) and (one+two)-body
(FULL) results are listed separately.  Note that the IA results depend on the
cutoff $\Lambda$ through the nuclear potentials.
Theoretical errors in the FULL results arise from the 
fitting procedure, and are due primarily to the experimental
error on GT$^{\rm EXP}$.  They are not indicated
when less than 0.1 sec$^{-1}$.  Electroweak radiative 
corrections have been included as estimated in Ref.~\cite{Cza07} 
for hydrogen and $^3$He---we have assumed
that those for deuterium are the same as for hydrogen.  The cutoff
dependence of the predictions is weak, at less than 1\% level, thus
suggesting that the mismatch between the chiral order of
the potentials and that of the currents may be of little numerical import.
If we also account for uncertainties in the electroweak radiative
corrections of the order of 0.4\%~\cite{Cza07}, we can conservatively
quote $\Gamma(^2{\rm H})=399\pm 3$ sec$^{-1}$ and 
$\Gamma(^3{\rm He})=1494\pm 21$ sec$^{-1}$.  These
predictions are in good agreement with available experimental data
(although those on $\Gamma(^2{\rm H})$~\cite{Gor04} have large errors),
as well as with results of recent theoretical studies~\cite{Mar11,And02,Gaz08}.
Finally, a comparison between the calculated and measured $\mu$--3
rates makes it possible to put a constraint on the 
induced pseudoscalar f.f.\ $G_{PS}(q^2)$ at $q_0^2=-0.954\, m_\mu^2$
relevant for the $\mu$--3 reaction.  By varying
$G_{PS}(q_0^2)$ so as to match the theoretical upper (lower) value
with the experimental lower (upper) value for the rate, we obtain
$G_{PS}(q_0^2)=8.2 \pm 0.7$, in good agreement with the $\chi$PT
prediction of $7.99 \pm 0.20$~\cite{GPth}.

\begin{widetext}
\begin{center}
\begin{table}[bthp]
\caption{Total rates for muon capture on deuteron $\Gamma(^2{\rm H})$
and $^3$He $\Gamma(^3{\rm He})$, in sec$^{-1}$, corresponding
to $\Lambda=500$ and 600 MeV.  The one-body (IA) and
(one+two)-body (FULL) contributions are listed, along with the individual
partial-wave contributions to $\Gamma(^2{\rm H})$.
Theoretical uncertainties in the FULL results, 
not reported when below 0.1 sec$^{-1}$,
are due to the fitting procedure.
}
\begin{tabular}{c|ccccccc|c}
\hline
& $^1S_0$ & $^3P_0$ & $^3P_1$ 
& $^3P_2$ & $^1D_2$ & $^3F_2$ & $\Gamma(^2{\rm H})$ 
& $\Gamma(^3{\rm He})$ \\
\hline
IA($\Lambda=500$ MeV) & 238.8 & 21.1 & 44.0 & 72.4 & 4.5 & 0.9 & 
381.7 & 1362 \\
IA($\Lambda=600$ MeV) & 238.7 & 20.9 & 43.8 & 72.0 & 4.5 & 0.9 & 
380.8 & 1360 \\
\hline
FULL($\Lambda=500$ MeV) & 254.4$\pm 0.9$ & 20.5 & 46.8 & 72.1 & 4.5 & 0.9 & 
399.2$\pm 0.9$ & 1488$\pm 9$ \\
FULL($\Lambda=600$ MeV) & 255.2$\pm 1.0$ & 20.3 & 46.6 & 71.6 & 4.5 & 0.9 & 
399.1$\pm 1.0$ & 1499$\pm 9$ \\
\hline
\end{tabular}
\label{tab:res}
\end{table}
\end{center}
\end{widetext}

The authors would like to thank P.\ Kammel for encouraging us to
carry out this study, and 
D.\ Gazit, P.\ Navr\'atil and S.\ Quaglioni
for useful discussions.  The work of R.S.\ is supported by the U.S.~Department
of Energy, Office of Nuclear Physics under contract DE-AC05-06OR23177.

\end{document}